\documentclass[notitlepage,report,nofootinbib,twocolumn]{revtex4-1}
\usepackage{graphicx}
\usepackage{booktabs}
\usepackage[breaklinks,hidelinks]{hyperref}
\usepackage{amssymb,amsmath,amsfonts}
\usepackage[utf8]{inputenc}
\usepackage[normalem]{ulem}
\usepackage{color}
\usepackage{xspace}

\newcommand{\ud}{\mathrm{d}}
\newcommand{\pt}{\ensuremath{p_{\mathrm{T}}}}
\newcommand{\mub}{\ensuremath{\mu_{B}}\xspace}
\newcommand{\dndy}{\ensuremath{\ud N/\ud y}\xspace}
\newcommand{\ee}{\ensuremath{\mathrm{e}^+\mathrm{e}^-}\xspace}
\newcommand{\jpsi}{\ensuremath{\mathrm{J}/\psi}\xspace}

\newcommand{\overbar}[1]{\mkern 3.5mu\overline{\mkern-3.5mu#1\mkern-3.5mu}\mkern 3.5mu}

\renewcommand\sout{\bgroup \color{blue} \ULdepth=-.5ex \ULset}
\begin{document}

\title{Decoding the phase structure of QCD via particle production at high energy}

\author{Anton Andronic}
\affiliation{Research Division and EMMI, GSI Helmholtzzentrum f\"ur Schwerionenforschung, 64291 Darmstadt, Germany}
\affiliation{Institut f\"ur Kernphysik, Universit\"at M\"unster, 48149  M\"unster, Germany}
\author{Peter Braun-Munzinger}
\affiliation{Research Division and EMMI, GSI Helmholtzzentrum f\"ur Schwerionenforschung, 64291 Darmstadt, Germany}
\affiliation{Physikalisches Institut, Universit\"at Heidelberg, 69120 Heidelberg, Germany}
\affiliation{Institute of Particle Physics and Key Laboratory of Quark and Lepton Physics (MOE), Central China Normal University, Wuhan 430079, China}
\author{Krzysztof  Redlich}
\affiliation{University of Wroc\l aw, Institute of Theoretical Physics, 50-204 Wroc\l aw, Poland}
\affiliation{Research Division and EMMI, GSI Helmholtzzentrum f\"ur Schwerionenforschung, 64291 Darmstadt, Germany}
\author{Johanna  Stachel}
\affiliation{Physikalisches Institut, University of Heidelberg, 69120 Heidelberg, Germany}

\begin{abstract}
\centerline{\bf Abstract}
\vskip 0.2cm

Recent studies based on non-perturbative lattice Monte-Carlo solutions of
Quantum Chromodynamics, the theory of strong interactions,  demonstrated
that at high temperature there is a phase change from confined
hadronic matter to a deconfined quark-gluon plasma where quarks and
gluons can travel distances largely exceeding the  size of hadrons. The
phase structure of such strongly interacting matter can be decoded via
analysis of particle abundances in high energy nuclear collisions within
the framework of the statistical hadronization approach. The results
imply quark-hadron duality at and experimental delineation of the
location of the phase boundary of strongly interacting matter.

\end{abstract}
\maketitle\let\endtitlepage\relax

\section*{} 

Atomic nuclei are bound by the strong force between their
constituents, protons and neutrons, or 'nucleons'. Although the
density in the center of a heavy nucleus is extremely large (about
$10^{14}$ times the density of water), the mean distance between
nucleons exceeds their diameter (the radius of the nucleon is about
0.88~fm = 0.88$\cdot 10^{-15}$ m, and the number density inside a nucleus is
$n_0 = 0.16/\rm{fm}^3$). Thus in terms of the filling fraction,
normal nuclear matter is dilute. If one compresses or heats such
matter in high energy nuclear collisions (see, e.g.,
\cite{Gyulassy:2004zy,BraunMunzinger:2007zz,Jacak:2012dx}) to even
higher densities and/or high temperatures (typically of order of $k_BT
\approx m_{\pi} c^2$ where $m_{\pi}$ is the mass of the lightest
hadron, the pion, and $k_B$ and $c$ are Boltzmann's constant and the
speed of light), one expects
\cite{Itoh:1970uw,Collins:1974ky,Cabibbo:1975ig,Chapline:1976gy} that
quarks, the building blocks of nucleons, are no longer confined but
can move over distances much larger than the size of the nucleon.
Such a 'deconfined' state of matter, named the Quark-Gluon Plasma
(QGP) \cite{Shuryak:1978ij}, is likely to have existed in the Early
Universe within the first microseconds after its creation in the Big
Bang \cite{Boyanovsky:2006bf}. One of the challenging questions in
modern nuclear physics is to identify the structure and phases of such
strongly interacting matter \cite{Rajagopal:2000wf}.

Evidence for the existence, in the laboratory, of the QGP was obtained
by studying collisions between heavy atomic nuclei (Au, Pb) at ultra-relativistic
energies. First relevant results came from experiments at the CERN
Super-Proton-Synchrotron (SPS) accelerator \cite{Heinz:2000bk}.  With
the start of operation of the Relativistic Heavy Ion Collider (RHIC)
at Brookhaven National Laboratory (BNL), experiments confirmed the
existence of this new state of matter  and found new observables providing further strong
evidence for QGP formation and expansion dynamics in the hot fireball
produced in high-energy nuclear collisions. Interesting and supporting
evidence was also obtained from experiments at the BNL Alternating
Gradient Synchrotron (AGS) through the discovery of collective
dynamics at high energy \cite{Barrette:1994xr}. For nuclear
collisions the center-of-mass energies per nucleon pair,
$\sqrt{s_{\rm NN}}$, covered by different accelerator facilities are:
\begin{enumerate}
\item BNL AGS, $\sqrt{s_{\rm NN}}$ = 2.7 - 4.8 GeV
\item CERN SPS, $\sqrt{s_{\rm NN}}$ = 6.2 - 17.3 GeV
\item BNL RHIC, $\sqrt{s_{\rm NN}}$  = 7.0 - 200 GeV
\item CERN LHC, $\sqrt{s_{\rm NN}}$  = 2.76 - 5.02 TeV
\end{enumerate}

Importantly, the results from RHIC showed that the QGP behaves more
like a nearly ideal, strongly interacting fluid rather than a weakly
interacting gas of quarks and gluons
\cite{Adams:2005dq,Arsene:2004fa,Adcox:2004mh,Back:2004je,Gyulassy:2004zy,Jacak:2012dx}. 
These results were confirmed and extended into hitherto
unexplored regions of phase space (in particular high transverse momenta) by experiments at the CERN Large Hadron Collider (LHC)
\cite{Muller:2012zq,Schukraft:2013wba,Braun-Munzinger:2015hba}. At LHC
energies the fireball formed in Pb-Pb collisions is so hot and dense
that quarks or gluons (partons) produced initially with energies of up
to a few hundred GeV lose a significant fraction of their energy while
traversing it.

The characterization of the QGP in terms of its equation of state
(EoS) expressing pressure as function of energy density, and of its
transport properties such as, e.g., its viscosity or diffusion
coefficients, as well as delineating the phases of strongly
interacting matter \cite{BraunMunzinger:2008tz} is a major ongoing
research effort
\cite{BraunMunzinger:2007zz,Jacak:2012dx,Schukraft:2013wba,Muller:2013dea,Satz:2013xja}.
However, it turned out that direct connections between the underlying
theory of strong interactions in the standard model of particle
physics, Quantum Chromodynamics (QCD) \cite{Patrignani:2016xqp} and the
experimental data are not readily established. This is since the
constituents of the QGP, the colored quarks and gluons, are not
observable as free particles, a fundamental property of QCD called
'confinement'. What is observable are colorless bound states of these
partons, resulting in mesons and baryons, generally referred to as
hadrons. Furthermore, the equations of QCD can only be solved
analytically in the high energy and short distance limit where
perturbative techniques can be used due to the asymptotic freedom
property of QCD \cite{Gross:1973id,Politzer:1973fx}. This is
unfortunately not possible for the QGP where typical distance scales
exceed the size of the largest atomic nuclei and the typical momentum
scale is low. The only presently known technique is to solve the QCD
equations numerically by discretization of the QCD Lagrangian
on a four-dimensional space-time lattice, and statistical evaluation via 
Monte Carlo methods, Lattice QCD (LQCD) \cite{Karsch:2001cy}.

Below we will discuss  how  the phase structure of strongly interacting 
matter described by LQCD can be decoded via analysis of particle production
in high energy nuclear collisions. This is achieved by making use of
the observed thermalization pattern of particle abundances within the
framework of the statistical hadronization approach at various collision
energies.

\section*{Connecting  hadronic states and QCD constituents}
From LQCD calculations, a deconfinement transition from matter
composed of hadronic constituents, hadronic matter, to a QGP was
indeed predicted (see Ref.~\cite{Karsch:2001cy} for an early review)
at an energy density of about 1 GeV/fm$^3$.
Besides deconfinement, there is  also a chiral symmetry restoration transition 
expected in high energy density matter \cite{Wilczek:2000ih,Bazavov:2011nk}.

  Due to the very small masses of the up and down quarks, the equations of QCD 
exhibit symmetries, called chiral symmetries, that allow separate 
transformations among the right-handed  quarks (with spin oriented in 
direction of momentum) and left-handed quarks.  Such  symmetries, however, 
are not manifest  in the observed strongly interacting particles, as they do 
not come in opposite parity pairs. Thus, chiral symmetry must be spontaneously 
broken at finite energy density. Consequently, QCD predicts the existence of
a chiral transition between a phase where chiral symmetry is broken, at low 
temperature and/or density, and a chirally-symmetric phase at high temperature 
and/or density. The connection between deconfinement and chiral transition is 
theoretically not fully understood.

It was demonstrated \cite{Aoki:2006we}, again using the methods of LQCD, that 
at zero baryo-chemical potential \mub the deconfinement transition is linked to 
the restoration of chiral symmetry and that it is of crossover type with a 
continuous, smooth but rapid increase of thermodynamic quantities in a narrow
region around the pseudo-critical temperature $T_c$. 
Using here and below units such that $\hbar = 1$, $k_B = 1$, and $c=1$, the
value of $T_c$ at vanishing \mub is currently calculated in LQCD to be 154$\pm 9$ MeV \cite{Bazavov:2014noa} and 156$\pm 9$ MeV \cite{Borsanyi:2010bp,Borsanyi:2013bia} using different fermion actions, resulting in excellent agreement.  
Recent LQCD results also quantify the small decrease of $T_c$ with increasing
\mub as long as $\mub < 3 T_c$. Within this parameter range the transition is 
still of crossover type.  
A fundamental question is the possible existence of a critical end
point, where a genuine second order chiral phase transition is
expected.  This is currently addressed both experimentally (see a
review in Ref.~\cite{Luo:2017faz}) and theoretically (see a review in
Ref.~\cite{Karsch:2013fga}) and is one of the outstanding problems
remaining to understand the phase structure of hot and dense QCD
matter.

These results do not shed light on the mechanism of the transition
from deconfinement to confinement. In fact, the crossover nature of
the chiral transition raises the question whether hadron production
from a deconfined medium also might happen over a wide range of
temperatures and how confinement can be implemented in a smooth
transition without leading to free quarks. A related question is the
possible survival of colorless bound states (hadrons) in a deconfined
medium. The present work attempts to shed light on some of these
questions by making contact with LQCD phenomenology and the by now
impressive body of results on hadron production in central
collisions between two heavy atomic nuclei at high energy.
Central collisions are nearly head-on collisions; centrality is calculated
in experiment matching measured particle multiplicity or energy to the
geometry of the collision, see details in ref.~\cite{Muller:2012zq}.

Towards that end we remark that the QCD Lagrange density is formulated
entirely in terms of the basic constituents of QCD, the quarks and gluons.
The masses of hadrons as colorless bound-states of quarks and gluons are well 
calculated within LQCD, showing remarkable agreement with experiment \cite{Durr:2008zz}. 
This confirms that chiral symmetry is broken in the QCD vacuum, reflected 
in the mass differences between parity partners as well as the existence of 
anomalously light pions as approximate Goldstone bosons associated with spontaneous symmetry breaking.

One of the consequences of confinement in QCD is that physical
observables require a representation in terms of hadronic
states. Indeed, as has been noted recently in the context of QCD
thermodynamics (see, e.g., \cite{Bazavov:2017dus} and refs. therein)
the corresponding partition function $Z$ can be very well approximated
within the framework of the hadron resonance gas, as long as the
temperature stays below $T_c$. To make this more transparent, we first
note that all thermodynamic variables like pressure $P$ and entropy
density $s$ can be expressed in terms of derivatives of logarithms of
$Z$. For the pressure, e.g., we obtain for a system with volume $V$ and
temperature $T$:

\begin{equation}
\frac{p}{T^4} = \frac{1}{T^3} \frac{ \partial \ln Z(V,T,\mu)}{\partial V} \; .
\label{pressure}
\end{equation}
The results of \cite{Andronic:2012ut} imply that, as long as
$T \lesssim T_c$,
\begin{multline}
\ln Z(T,V,\mu) \approx
\sum_{i\in\;mesons}\hspace{-3mm}
\ln{\cal Z}^{M}_{M_i}(T,V,\mu_Q,\mu_S) +\\
\sum_{i\in\;baryons}\hspace{-3mm} \ln{\cal Z}^{B}_{M_i}(T,V,\mub,\mu_Q,\mu_S)\; ,
\label{ZHRG}
\end{multline}
where the partition function of the hadron resonance gas model is
expressed in mesonic and baryonic components, where $M_i$ is the mass of a given hadron. The chemical potential
$\mu$ reflects then the baryonic, electric charge, and strangeness components
$\mu = (\mub,\mu_Q,\mu_S)$.

To make this connection quantitative, detailed investigations have
recently been made on the contribution of mesons and baryons to the
total pressure of the matter. In particular, in
\cite{Karsch:2013naa,Bazavov:2017dus} and refs. therein, the equation
of state and different fluctuation observables are evaluated in the
hadronic sector via the hadron resonance gas and compared to predictions 
from LQCD. Very good agreement is obtained for temperatures up to very close 
to $T_c$, lending further support to the hadron-parton duality described by 
Eq.~\ref{ZHRG}.

The partition function of the hadron resonance gas in Eq.~\ref{ZHRG}
is evaluated as a mixture of ideal gases of all stable hadrons and
resonances. In the spirit of the S-matrix formalism
\cite{Dashen:1969ep}, which provides a consistent theoretical
framework to implement interactions in a dilute many-body system in
equilibrium, the presence of resonances corresponds to attractive
interactions among hadrons. This is generally a good approximation
since for temperatures considered here ($T < 165$ MeV) the total
particle density is low, $n < 0.5/\rm{fm}^3$.

Sometimes, additional repulsive interactions are modelled with an 'excluded
volume' prescription, see, e.g., \cite{Andronic:2012ut} and
refs. there, which is inherently a low density approach.
For weak repulsion, implying excluded volume radii $r_0 < 0.3$ fm,
the effect of the correction is mainly to decrease particle densities,
while the important thermal parameters $T$ and \mub are little
affected. Strong repulsion cannot be modelled that way: significantly
larger $r_0$ values lead to, among others, unphysical (superluminous)
equations of state, in contra-distinction to results from LQCD.
In the following we use $r_0 = 0.3$ fm both for mesons and
baryons. All results on thermal parameters described below are
unchanged from what is obtained in the non-interacting limit except
for the overall particle density which is reduced by up to 25\%.

Over the course of the last 20 years it has become apparent
\cite{Cleymans:1992zc,BraunMunzinger:1994xr,BraunMunzinger:2003zd,BraunMunzinger:2001ip,Letessier:2005qe,Stachel:2013zma}
that the yields of all hadrons produced in central collisions can be very well
described by computing particle densities from the above hadronic partition
function.
To obtain particle yields at a particular temperature 
$T_{CF}$ and $\mub$ one multiplies the so obtained thermal densities with the
fireball volume $V$. In practice, $T_{CF}$, $\mub$, and $V$, the
parameters at 'chemical freeze-out' from which point on all hadron
yields are frozen, are determined from a fit to the experimental data.
Note that $V$ is actually the volume corresponding to a slice of one
unit of rapidity, centered at mid-rapidity.
Experimentally, the rapidity density \dndy of a hadron is obtained by integration of its momentum space distribution over the momentum component transverse to the beam direction. In general we take these yields at mid-rapidity, i.e. $y = 0$, where the center of mass of the colliding system is at rest.

As will be discussed below, this 'statistical hadronization approach' provides,
via Eq.~\ref{pressure} and Eq.~\ref{ZHRG}, a link between data on hadron
production in ultra-relativistic nuclear collisions and the QCD
partition function. This link opens an avenue to shed light on the QCD
phase diagram. The possibility of such a connection was surmised early
on \cite{Cabibbo:1975ig,Hagedorn:1984hz} and various aspects of it
were discussed more recently
\cite{Cleymans:1998fq,Stock:1999hm,BraunMunzinger:2001mh,BraunMunzinger:2003zz,Letessier:2005qe,Andronic:2008gu,Floerchinger:2012xd,Bazavov:2012vg}.

The full power of this link, however, becomes apparent only with the
recent precision data from the LHC. Below we will discuss the
accuracy which can be achieved in the description of hadron production
using the parton-hadron duality concept described above. We will first
focus on hadrons containing only light quarks with flavors up, down,
and strange (u,d,s) and place emphasis on LHC data. Those show matter 
and anti-matter production in equal amounts, therefore indicating that 
\mub is very close to zero. It is in this energy region where the LQCD
approach can be applied essentially without approximations using
present computer technology. 
We then explore the lower energy region (500 GeV $ >\sqrt{s_{\rm NN}} > 15$ GeV) 
and show that, for the first time, consistent information on the QCD phase
 diagram for $0 < \mub < 300$ MeV can be achieved by
quantitatively comparing LQCD predictions for finite \mub with
results from statistical hadronization analysis of hadron production
data. In the last section we will discuss how the statistical
hadronization approach can be extended to include heavy (charm c and
bottom b) quarks. We further discuss how the recent LHC data can
provide information on the hadronization of these heavy quarks during
the expansion and cooling of the QGP formed in such high-energy
central nuclear collisions.

\section*{Statistical Hadronization of Light Quarks}
The description of particle production in nucleus-nucleus collisions
in the framework of the statistical hadronization approach is particulary
transparent at the LHC energy where the chemical freeze-out is
quantified, essentially, by the temperature $T_{CF}$ and the volume $V$ of 
the produced fireball. 

The parameters of the statistical hadronization approach are obtained with 
considerable precision by comparing to the yields of particles measured by 
the ALICE collaboration
\cite{Abelev:2013vea,Abelev:2013xaa,Abelev:2013zaa,Abelev:2014uua,Adam:2015yta,Adam:2015vda,Acharya:2017bso}.
To match the measurement, the calculations include all contributions
from the strong and electromagnetic decays of high-mass resonances.
For $\pi^\pm$, $K^\pm$, and $K^0$ mesons, the contributions from heavy
flavor hadron decays are also included.  The measurement uncertainty,
$\sigma$, is accounted for as the quadratic sum of statistical and
systematic uncertainties, see below.

\begin{figure}[hbt]
\hspace{-.2cm}\includegraphics[width=.49\textwidth]{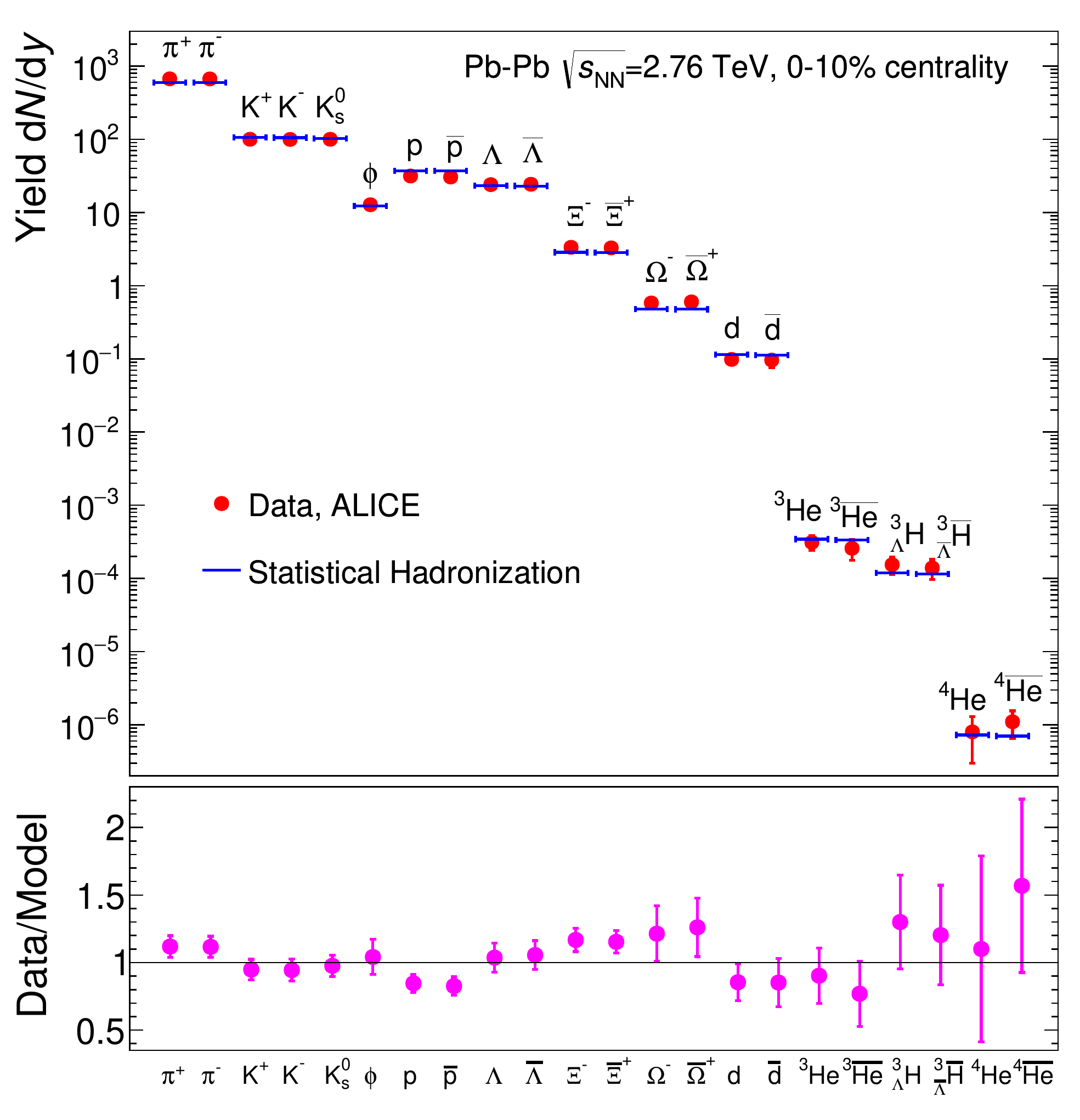}
\caption{Hadron abundances and statistical hadronization model predictions. Here \dndy values for different hadrons measured at midrapidity are compared with the statistical hadronization analysis.  The data are from the ALICE collaboration for central Pb--Pb collisions at the LHC. The lower panel shows the ratio of data and statistical hadronization predictions with uncertainties determined only from the data.
}
\label{fig:Fit}
\end{figure}

For the most-central Pb--Pb collisions, the best description of the
ALICE data on yields of particles in one unit of rapidity at
midrapidity, is obtained with $T_{CF}=156.5\pm 1.5$ MeV, $\mu_B=0.7\pm
3.8$ MeV, and $V=5280\pm 410$ fm$^3$. This result is an update of the
previous analysis from Ref. \cite{Stachel:2013zma} using an extended and
final set of data. The standard deviations quoted here are
exclusively due to experimental uncertainties and do not reflect the
systematic uncertainties connected with the model implementation, as
discussed below.

Remarkably, the value of the chemical freeze-out temperature
$T_{CF}=156.5 \pm1.5$ MeV and the pseudo-critical temperature,
$T_c=154 \pm 9$ MeV obtained in LQCD, agree within errors. This
implies that chemical freeze-out takes place close to hadronization of
the QGP, lending also support to the hadron-parton duality described
by Eq.~\ref{ZHRG}.

A comparison of the statistical hadronization results obtained with the thermal
parameters discussed above and the ALICE data for particle yields is shown in
Fig.~\ref{fig:Fit}.
Impressive overall agreement is obtained between the measured particle
yields and the statistical hadronization analysis.  The agreement spans nine
orders of magnitude in abundance values, encompasses strange and non-strange
mesons, baryons including strange and multiply-strange hyperons as
well as light nuclei and hypernuclei and their anti-particles.  A very
small value for the baryo-chemical potential $\mu_B = 0.7 \pm 3.8$ MeV,
consistent with zero, is obtained, as expected by the observation of
equal production of matter and antimatter at the LHC
\cite{Abelev:2012wca}.

The largest difference between data and calculations is observed for
proton and antiproton yields, where a deviation of 2.7$\sigma$ is
obtained. This difference is connected with an unexpected and puzzling
centrality dependence of the ratio $(\mathrm{p}+\bar{\mathrm{p}})/(\pi^++\pi^-)$
\cite{Abelev:2013vea}, see, in particular, Fig. 9 of this
reference. As discussed below, the other ratios (hadrons/pions)
increase towards more central collisions until a plateau (the
grand-canonical limit) is reached. The peculiar behavior of the
$(\mathrm{p}+\bar{\mathrm{p}})/(\pi^++\pi^-)$ at LHC energy is currently not
understood. Arguments that this might be connected to annihilation of
baryons in the hadronic phase after chemical freeze-out
\cite{Becattini:2014hla} are not supported by the results of recent
measurements of the relative yields of strange baryons to pions \cite{ALICE:2017jyt}.

\begin{figure}[hbt]
\hspace{-.4cm}\includegraphics[width=.5\textwidth]{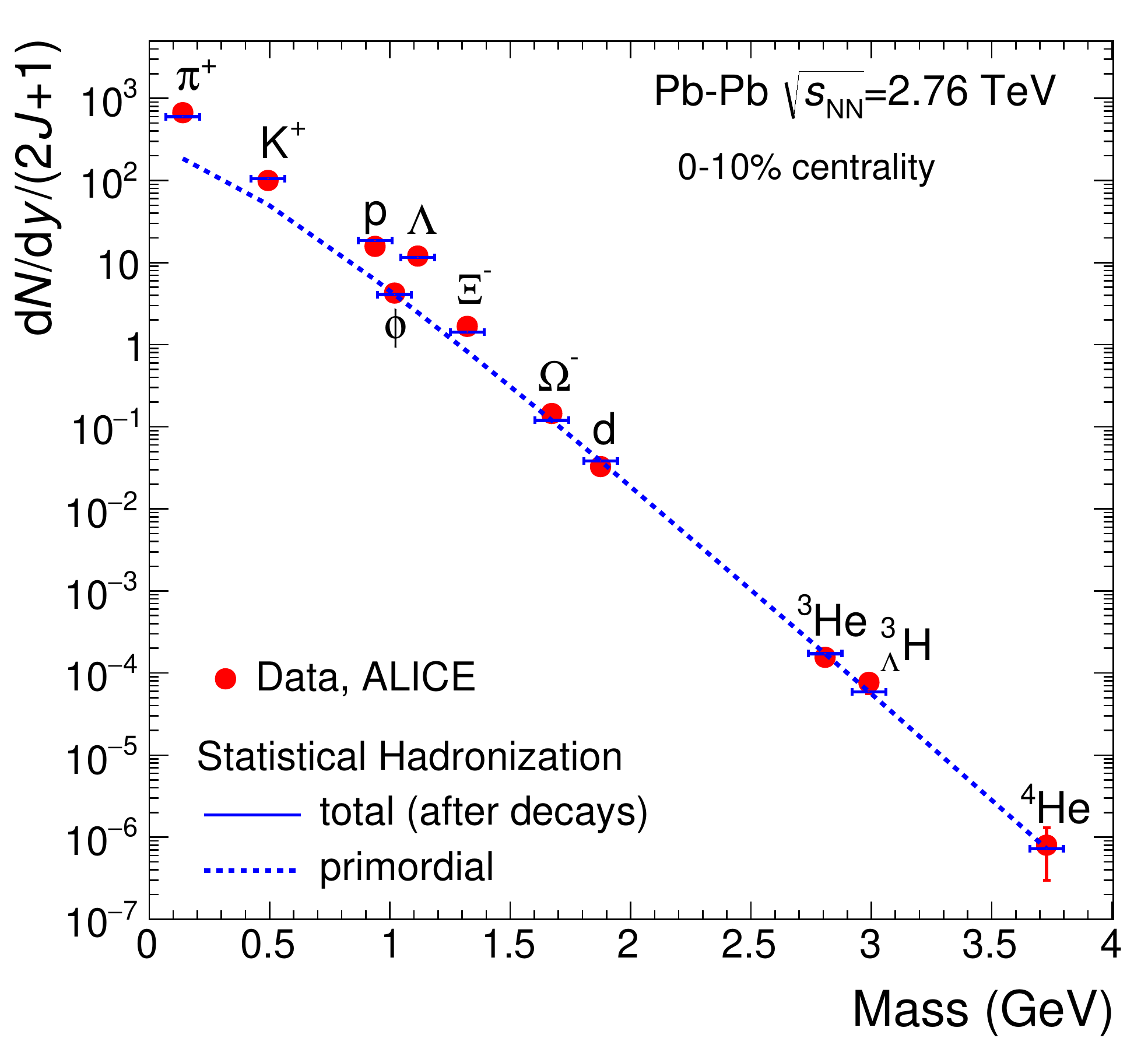}
\caption{Mass dependence of hadron yields compared with predictions of the statistical hadronization model.
  Only particles, no anti-particles, are included and the yields are divided
  by the spin degeneracy factor ($2J+1$). Data are from the ALICE
  collaboration for central Pb--Pb collisions at the LHC.
  For the statistical hadronization approach, plotted are the ``total''
  yields, including all contributions from high-mass resonances (for
  the $\Lambda$ hyperon, the contribution from the electromagnetic
  decay $\Sigma^0\rightarrow\Lambda\gamma$, which cannot be resolved
  experimentally, is also included), and the (``primordial'') yields
  prior to strong and electromagnetic decays. For more details see
  text.}
\label{fig:LQCD_data}
\end{figure}

A further consequence of the vanishing baryo-chemical potential is
that also the strangeness chemical potential $\mu_S$ vanishes. This implies
that the strangeness quantum number plays no role anymore for the
particle production. In the fireball the yield of strange mesons and
(multi-)strange baryons is exclusively determined by their mass $M$
and spin degeneracy $(2J + 1)$ in addition to the temperature $T$.

The thermal origin of all particles including light nuclei and
anti-nuclei is particularly transparent when inspecting the change of
their yields with particle mass. This is shown in
Fig.~\ref{fig:LQCD_data} where the measured yields, normalized to the
spin degeneracy, are plotted as a function of the mass $M$. This
demonstrates explicitly that the normalized yields exclusively depend
on $M$ and $T$. For heavy particles ($M \gg T$) without resonance
decay contributions their normalized yield simply scales with mass as
$M^{3/2} \exp{(-M/T)}$, illustrated by the nearly linear dependence
observed in the logarithmic representation of
Fig.~\ref{fig:LQCD_data}. We note that, for the subset of light
nuclei, the statistical hadronization predictions are not
affected by resonance decays. For these nuclei, a small variation in
temperature leads to a large variation of the yield, resulting in a
relatively precise determination of the freeze-out temperature
$T_{nuclei} = 159 \pm 5$ MeV, well consistent with the value of
$T_{CF}$ extracted above.

The incomplete knowledge of the structure and decay probabilities of
heavy mesonic and baryonic resonances discussed above leads to
systematic uncertainties in the statistical hadronization
approach. We note, from Fig.~\ref{fig:LQCD_data}, that the yields of
the measured lightest mesons and baryons, ($\pi, K,p,\Lambda$) are
substantially increased relative to their primordial thermal
production by such decay contributions. For pions, e.g., the resonance
decay contribution amounts to 70\%. For resonance masses larger than
1.5 GeV the individual states start to strongly overlap
\cite{Patrignani:2016xqp}. Consequently, neither their number density nor
their decay probabilities can be determined well. Indeed, recent LQCD
results indicate that there are missing resonances compared to what is
listed in \cite{Patrignani:2016xqp}. The resulting theoretical
uncertainties are difficult to estimate but are expected to be small
since $T_{CF}$ is very small compared to their mass. A conservative
estimate is that the resulting systematic uncertainty in $T_{CF}$ is
at most $3\%$. This is consistent with the determination of
$T_{CF}$ using only particles whose yields are not influenced by
resonance decays, see above. Until now none of these systematic
uncertainties are taken into account in the statistical hadronization
analysis described here.

The rapidity densities of light (anti)-nuclei and hypernuclei were
actually predicted \cite{Andronic:2010qu}, based on the systematics of
hadron production at lower energies. It is nevertheless remarkable
that such loosely bound objects (the deuteron binding energy is 2.2
MeV, much less than $T_{nuclei} \approx  159$ or $T_{CF} \approx T_c  \approx 155$ MeV)
are produced with temperatures very close to that of the phase
boundary at LHC energy, implying any further evolution of the fireball
has to be close to isentropic. For the (anti-)hypertriton the
situation is even more dramatic: this object consists of a bound state
of (p, n, $\Lambda$), with a value of only $130 \pm 30$ keV for the
energy needed to remove the $\Lambda$ from it. This implies that the
$\Lambda$ particle is very weakly bound to a deuteron, resulting in a value
for the root-mean-square size for this bound state of close to 10
fm, about the same size as that of the fireball formed in the Pb--Pb
collision.

The detailed production mechanism for loosely bound states remains an
open question. One, admittedly speculative, possibility is that such
objects, at QGP hadronization, are produced as compact, colorless
droplets of quark matter with quantum numbers of the final state
hadrons. The concept of possible excitations of nuclear matter
  into colorless quark droplets was considered already in
  \cite{Chapline:1978kg}. In our context, these states should have a
lifetime of 5 fm or longer, excitation energies of 40 MeV or less, for
evolution into the final state hadrons which are measured in the
detector. Since by construction they are initially compact they would
survive also a possible short-lived hadronic phase after
hadronization. This would be a natural explanation for the striking
observation of the thermal pattern for these nuclear bound states
emerging from Figs.~\ref{fig:Fit} and ~\ref{fig:LQCD_data}. Note that
the observed thermal nature of their production yields is very
difficult to reconcile with the assumption that these states are
formed by coalescence of baryons, where the yield is proportional to a
coalescence factor introduced as the square of the nuclear wave
function, which actually differs strongly for the various nuclei
\cite{Csernai:1986qf,Hirenzaki:1992gx}. For a recent discussion of the
application of coalescence models to production of loosely bound
states, see ~\cite{Cho:2017dcy}.

One might argue that composite particles such as light nuclei and
hypernuclei should not be included in the hadronic partition function
described in Eq.~\ref{ZHRG}. We note, however, that all nuclei,
including light, loosely bound states, should result from the
interaction of the fundamental QCD constituents. This is confirmed by
recent LQCD calculations, see \cite{Beane:2012vq}.

\begin{figure}[htb]
\includegraphics[width=.46\textwidth]{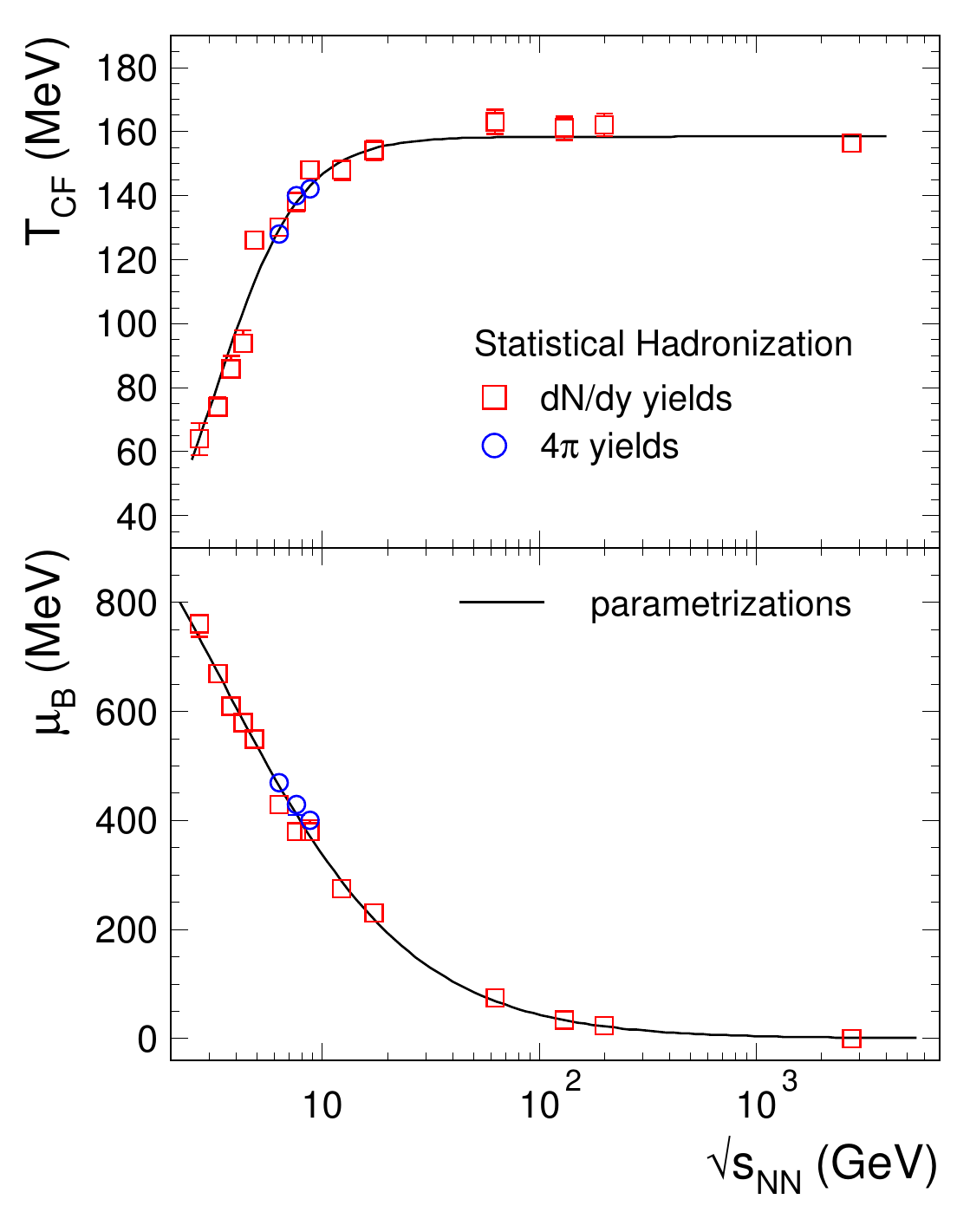}
\caption{Energy dependence of chemical freeze-out parameters $T_{CF}$
  and \mub. The results are obtained from the statistical hadronization
  analysis of hadron yields (at midrapidity, \dndy, and in full phase space, 
  $4\pi$) for central collisions at different energies.
  The parametrizations shown are:
$T_{CF}={T_{CF}^{lim}}/(1+\exp(2.60-\ln(\sqrt{s_{\rm NN}})/0.45))$,
$\mu_B ={a}/(1+0.288\sqrt{s_{\rm NN}})$, with $\sqrt{s_{\rm NN}}$ in GeV and 
$T_{CF}^{lim}=158.4$ MeV and $a=1307.5$ MeV; 
the uncertainty of the 'limiting temperature', $T_{CF}^{lim}$, determined from 
the fit of the 5 points for the highest energies, is 1.4 MeV.}
\label{fig:edep}
\end{figure}

The thermal nature of particle production in ultra-relativistic
nuclear collisions has been experimentally verified not only at LHC
energy, but also at the lower energies of the RHIC, SPS and AGS
accelerators. The essential difference is that, at these lower
energies, the matter-antimatter symmetry observed at the LHC is
lifted, implying non-vanishing values of the chemical
potentials. Furthermore, in central collisions at energies below
$\sqrt{s_{\rm NN}} \approx 6 $ GeV the cross section for the production of
strange hadrons decreases rapidly, with the result that the average
strange hadron yields per collision can be significantly below
unity. In this situation, one needs to implement exact strangeness
conservation in the statistical sum in Eq.~\ref{ZHRG} and apply the
canonical ensemble for the conservation laws
\cite{Hagedorn:1984uy,Hamieh:2000tk}. Similar considerations apply for
the description of particle yields in peripheral nuclear and
elementary collisions.  An interesting consequence of exact
strangeness conservation is a suppression of strange particle yields
when going from central to peripheral nucleus-nucleus collisions or
from high multiplicity to low multiplicity events in proton-proton or
proton-nucleus collisions. In all cases the suppression is further
enhanced with increasing strangeness content of hadron. Sometimes,
additional fugacity parameters $g_f$ are introduced to account for
possible non-equilibrium effects  of strange and heavy flavor hadrons \cite{Letessier:2005qe,BraunMunzinger:2000px}. These parameters  modify the thermal 
yields of particles by factors $g_f^{n_f}$, where the power $n_f$ denotes
the number of strange or heavy quarks and anti-quarks in the hadron.

\begin{figure}[htb]
\hspace{-.4cm}\includegraphics[width=.5\textwidth]{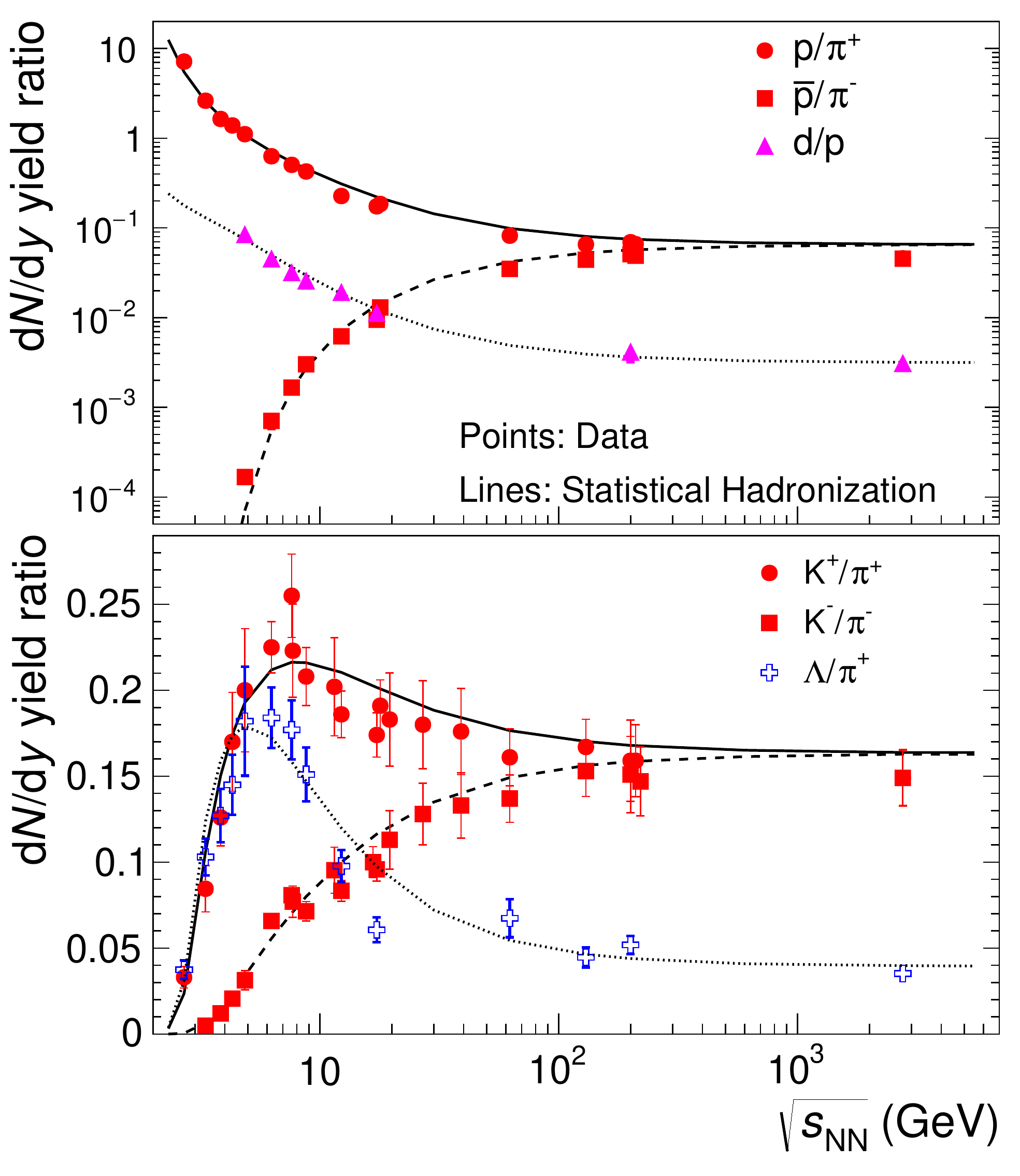}
\caption{Collision energy dependence of the relative abundance of
  several hadron species. The data (symbols) are compiled in \cite{Andronic:2014zha,Adamczyk:2017iwn} and are compared to statistical hadronization calculations for the smooth parametrization of $T_{CF}$ and $\mu_B$ as a function of energy shown in Fig.~\ref{fig:edep}.
}
\label{fig:ratios}
\end{figure}

Experimental consequences of canonical thermodynamics and strangeness
conservation laws have been first seen at SPS energy
\cite{Antinori:2004ee}. All above predictions are qualitatively
confirmed by the striking new results from high multiplicity
proton-proton and p-Pb collisions from the ALICE collaboration at LHC
energy \cite{ALICE:2017jyt}. The data also explicitly exhibit the
plateau in strangeness production when reaching Pb-Pb collisions which
is expected when the grand-canonical region is reached, further
buttressing the thermal analysis discussed above.

An intriguing observation, first made in \cite{Becattini:1995if}, is that the
overall features of hadron production in \ee annihilations resemble that
expected from a thermal ensemble with temperature $T \approx 160$ MeV, once
exact quantum number conservation is taken into account. In these collisions, quark-antiquark pairs are produced with production yields that are not thermal
but are well explained by the electro-weak standard model, see, e.g., Table II in \cite{Becattini:2008tx}.
Hadrons from these quark pairs (and sometimes gluons) appear as jets in the data. The underlying hadronisation
process can be well described using statistical hadronisation model
ideas \cite{Becattini:2008tx,Andronic:2008ev}. These studies  revealed
further that strangeness production deviates significantly from a pure thermal
production model and that the quantitative description  of the measured yields
is rather poor. Nevertheless, recognizable thermal features in \ee collisions,
where equilibration should be absent, may be a consequence of the generic nature
of hadronization in strong interactions.

From a statistical hadronization analysis of all measured hadron
yields at various beam energies the detailed energy dependence of the
thermal parameters $T_{CF}$ and \mub has been determined
\cite{BraunMunzinger:1994xr,Cleymans:1998yb,BraunMunzinger:1999qy,Andronic:2008gu,Manninen:2008mg,Abelev:2009bw,Adamczyk:2017iwn}.
While \mub decreases smoothly with increasing energy, the dependence
of $T_{CF}$ on energy exhibits a striking feature which is illustrated
in Fig.~\ref{fig:edep}: $T_{CF}$ increases with increasing energy
(decreasing \mub) from about 50 MeV to about 160 MeV, where it exhibits a
saturation for $\sqrt{s_{\rm NN}} > 20$ GeV.  The slight increase of this
value compared to $T_{CF} = 156.5$ MeV obtained at LHC energy is due
to the inclusion of points from data at RHIC energies, the details of
this small difference are currently not fully understood.

The saturation of $T_{CF}$ observed in Fig.~\ref{fig:edep} lends
support to the earlier proposal
\cite{BraunMunzinger:1998cg,Stock:1999hm,BraunMunzinger:2003zz} that,
at least at high energies, the chemical freeze-out temperature is very
close to the QCD hadronization temperature \cite{Andronic:2008gu},
implying a direct connection between data from relativistic nuclear
collisions and the QCD phase boundary. This is in accord with the
earlier prediction, already more than 50 years ago, by Hagedorn
\cite{Hagedorn:1965st,Hagedorn:1985js} that hadronic matter cannot be
heated beyond this limit. Whether there is, at the lower energies, a
critical end-point \cite{Stephanov:1998dy} in the QCD phase diagram is 
currently at the focus of intense theoretical \cite{Braun-Munzinger:2015hba} 
and experimental effort \cite{Adamczyk:2017iwn}.

To illustrate how well the thermal description of particle production
in central nuclear collisions works we show, in Fig.~\ref{fig:ratios},
the energy dependence (excitation function) of the relative abundance
of several hadron species along with the prediction using the
statistical hadronization approach and the smooth evolution of the
parameters (see above). Because of the interplay between the energy
dependence of $T_{CF}$ and $\mu_B$ there are 
characteristic features in these excitation functions. In particular, 
maxima appear at slightly different c.m. energies in the $K^+/\pi^+$
and $\Lambda/\pi^+$ ratios while corresponding anti-particle ratios
exhibit a smooth behavior \cite{BraunMunzinger:2001as}. 

In the statistical approach in  Eq.~\ref{ZHRG} and in the Boltzmann 
approximation, the density $n(\mu_B,T)$ of hadrons carrying baryon number $B$
scales with the chemical potential as  $n(\mu_B,T)\propto \exp(B\mu_B/T)$. 
Consequently, the ratios $p/\pi^+$ and $d/p$ scale as $\exp(\mu_B/T)$, whereas
the corresponding anti-particle ratios scale as $\exp(-\mu_B/T)$. 
From Fig.~\ref{fig:edep}, it is apparent that $\mu_B/T_{CF}$ decreases with 
collision energy, accounting for the basic features of 
particle ratios in the upper part of  Fig.~\ref{fig:ratios}. 
On the other hand, strangeness conservation unambigously connects, for every 
$T$ value, the strangeness- and baryo-chemical potentials, $\mu_S=\mu_S(\mu_B)$.
As a consequence, the yields of $K^+$ and $K^-$ increase and, respectively, decrease with $\mu_B/T$. At higher energies, where $T$ and hence pion densities saturate, the $\Lambda/\pi^+$  and $K^+/\pi^+$ ratios are decreasing with energy (see lower part of Fig.~\ref{fig:edep}).

 We further note that, for energies beyond that of the LHC, the thermal 
parameter $T_{CF}$ is determined by the QCD pseudo-critical temperature and 
the value of $\mu_B$ vanishes. Combined with the energy dependence of overall
particle production \cite{Adam:2015ptt} in central Pb-Pb collisions
this implies that the statistical hadronization model prediction of
particle yields at any energy, including those at the possible Future Circular
Collider (FCC) \cite{Dainese:2016gch} or in ultra-high energy cosmic ray
collisions \cite{Kampert:2016sqd}, can be made with an estimated
precision of better than 15\%.

\begin{figure}[htb]
\includegraphics[width=.5\textwidth]{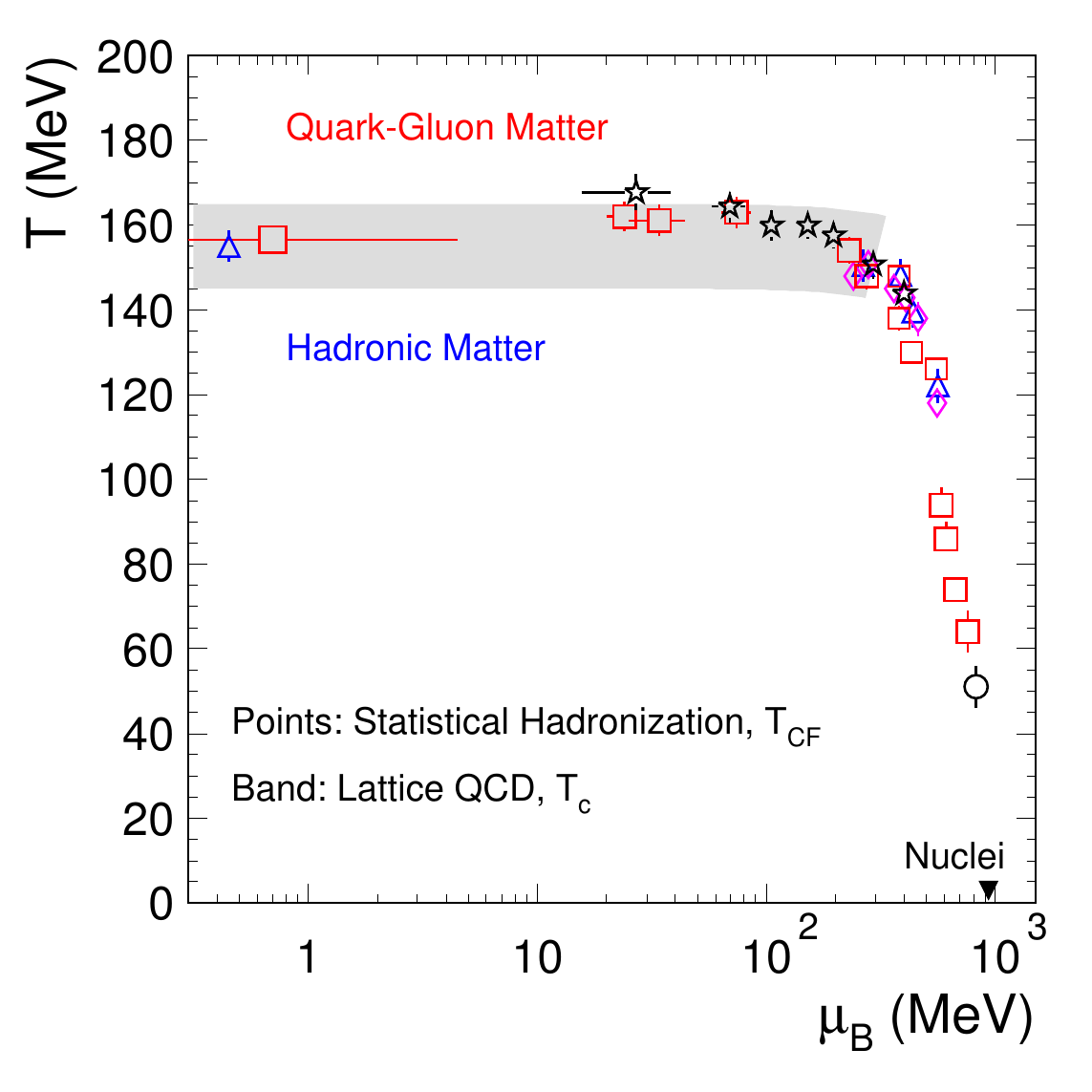}
\caption{Phenomenological phase diagram of strongly interacting matter
  constructed from chemical freeze-out points resulting from
  statistical hadronization analysis of hadron yields for central collisions at different energies.
 The freeze-out points extracted from experimental data sets in our own analysis (squares) and other similar analyses \cite{Cleymans:1998yb,Vovchenko:2015idt,Becattini:2016xct,Adamczyk:2017iwn} are compared
 to predictions from LQCD \cite{Bazavov:2014noa,Borsanyi:2012cr} shown
 as a band.  The inverted triangle marks the value for ground state
 nuclear matter (atomic nuclei).}
\label{fig:t-mu}
\end{figure}

Since the statistical hadronization analysis at each measured energy yields 
a pair of ($T_{CF}$,\mub) values, these points can be used to construct a $T$ 
vs. \mub diagram, describing phenomenological constraints on the phase boundary
between hadronic matter and the QGP, see Fig.~\ref{fig:t-mu}.  
Note that the points at low temperature seem to converge towards the value 
for ground state nuclear matter ($\mub = 931$ MeV). 
As argued in \cite{Floerchinger:2012xd} this limit is not 
necessarily connected to a phase transition.  While the situation at low 
temperatures and collision energies is complex and at present cannot be 
investigated with first-principle calculations, the high temperature, 
high collision energy limit allows a quantitative interpretation in terms of 
fundamental QCD predictions.

The connection between LQCD predictions and experimental chemical freeze-out
points is made quantitative in Fig.~\ref{fig:t-mu}. We take here
recent results for the QCD phase boundary from the two leading LQCD
groups \cite{Bazavov:2014noa,Borsanyi:2012cr}, represented by the band
in Fig.~\ref{fig:t-mu}. As can be seen, the LQCD values follow the
measured \mub dependence of the chemical freeze-out temperature very
closely, demonstrating that with relativistic nuclear collisions one
can directly probe the QCD phase boundary between hadronic matter and
the QGP.  The above results imply that the pseudo-critical temperature
of the QCD phase boundary at $\mu_B = 0$ as well as its $\mu_B$
dependence up to $\mu_B \leq 300$ MeV have been determined
experimentally. There is indirect but strong evidence from measurements of the initial energy density as well as from hydrodynamical analysis of transverse momentum spectra and from the analysis of jet quenching data that initial temperatures of the fireball formed in the collision are substantially higher than the values at the phase boundary, reaching 300-600 MeV at RHIC and LHC energies \cite{Adare:2008ab,Adam:2015lda}. 

We close this section by remarking that the present approach can be extended 
beyond hadron yields to higher moments of event-by-event particle distributions.
While precision predictions from LQCD exist for higher moments and 
susceptibilities, especially at LHC energies where \mub is close to zero, 
see,e.g. \cite{Karsch:2013naa,Bazavov:2017dus}, there are formidable challenges
to experimentally determine such higher moments with accuracy. Pioneering 
experiments where performed at the RHIC accelerator with intriguing but not yet
fully conclusive results, for a recent review see \cite{Luo:2017faz}.  
Analyses of higher moments are, therefore, not considered in the present paper.
Recently, however, variances of strangeness and net-baryon number fluctuations 
were reconstructed \cite{Braun-Munzinger:2014lba} from hadron yields measured 
in Pb-Pb collisions at mid-rapidity by the CERN ALICE collaboration. 
The so determined second moments are in impressive 
agreement with LQCD predictions obtained at the chiral crossover 
pseudo-critical temperature $T_c \approx 154$ MeV.
Furthermore, an interesting strategy was proposed to compare directly  
experimental data on moments of net-charge fluctuations with LQCD results
to identify freeze-out parameters in heavy ion collisions \cite{Karsch:2012wm}.
Within still large systematic uncertainties the extracted freeze-out parameters
based on 2nd order cumulants are well consistent with statistical hadronization  \cite{Borsanyi:2014ewa,Adare:2015aqk}.
 While no formal proof of the above discussed quark-hadron duality
near the chiral crossover temperature exists, the empirical evidence
has recently clearly been strengthened.

\section*{Statistical Hadronization of Heavy Quarks} 
An interesting question is whether the production of hadrons with
heavy quarks can be described with similar statistical hadronization
concepts as developed and used in the previous section for light
quarks. We note first that the masses of charm and beauty quarks, $m_c
= 1.28$ GeV and $m_b = 4.18$ GeV are much larger than the
characteristic scale of QCD, $\Lambda_{QCD} = 332$ MeV for three quark
flavors, in the $\overbar{\mathrm{MS}}$ scheme \cite{Patrignani:2016xqp}. 
Both masses are also sufficiently larger than the pseudo-critical
temperature $T_c$ introduced above, such that thermal production of
charm and, in particular, beauty quarks is strongly Boltzmann
suppressed. This is also borne out by quantitative calculations
\cite{BraunMunzinger:2000dv,Zhang:2007dm,Zhou:2016wbo}. On the other
hand we expect, in particular at LHC energies, copious production
of heavy quarks in relativistic nuclear collisions through hard
scattering processes which, in view of the large quark masses, can be
well described using QCD perturbation theory \cite{Cacciari:2012ny}.
Consequently, the charm and beauty content of the fireball formed in
the nucleus-nucleus collision is determined by initial hard
scattering. Furthermore, annihilation of c- and b-quarks is very small
implying that their numbers are closely preserved during the fireball
evolution \cite{Andronic:2006ky}.

The produced charm quarks will, therefore, not resemble a chemical
equilibrium population for the temperature $T$. However, what is
needed for the thermal description proposed below is that the heavy
quarks produced in the collision reach a sufficient degree of thermal
equilibrium through scattering with the partons of the hot medium.
Indeed, the energy loss suffered by energetic heavy quarks in the QGP
is indicative of their ``strong coupling'' with the medium, dominated
by light quarks and gluons.  
The measurements at the LHC \cite{ALICE:2012ab,Abelev:2013lca} and 
RHIC \cite{Adamczyk:2014uip} of the energy loss and hydrodynamic flow of 
D mesons  demonstrate this quantitatively.

Among the various suggested probes of deconfinement, charmonium (the
bound states of $c\bar{c}$) plays a distinctive role. The \jpsi
meson is the first hadron for which a clear mechanism of suppression
(melting) in the QGP was proposed early on, based on the color
analogue of Debye screening \cite{Matsui:1986dk}.  This concept for
charmonium suppression was tested experimentally at the SPS
accelerator but led, with the turn-on of the RHIC facility, to a
number of puzzling results. In particular, the observed rapidity and
energy dependence of the suppression ran counter to the theoretical
predictions \cite{Vogt:1999cu}.

However, still before publication of the first RHIC data, a novel quarkonium 
production mechanism, based on statistical hadronization, was proposed
\cite{BraunMunzinger:2000px} which contained a natural explanation for
the later observations at RHIC and LHC energy.  The basic concept
underlying this statistical hadronization approach
\cite{BraunMunzinger:2000px} is that charm quarks are produced in
initial hard collisions, subsequently thermalize in the QGP and are
"distributed" into hadrons at the phase boundary, i.e. at chemical
freeze-out, with thermal weights as discussed above for the light
quarks, \cite{Andronic:2006ky,BraunMunzinger:2000px,BraunMunzinger:2009ih,Andronic:2011yq}.  An alternative mechanism
for the (re)combination of charm and anti-charm quarks into charmonium
in a QGP \cite{Thews:2000rj} was proposed based on kinetic theory.
For further developments see
\cite{Liu:2009nb,Grandchamp:2003uw,Emerick:2011xu,Zhou:2014kka}.

In the statistical hadronization approach, the absence of chemical
equilibrium for heavy quarks is accounted for by introducing a
fugacity $g_c$. The parameter $g_c$ is obtained from the balance equation
\cite{BraunMunzinger:2000px} which accounts for the distribution of
all initially produced heavy quarks into hadrons at the phase
boundary, with a thermal weight constrained by exact charm
conservation. With the above approach the knowledge of the heavy quark
production cross section along with the thermal parameters obtained
from the analysis of the yields of hadrons composed of light quarks,
see previous section, is sufficient to determine the yield of hadrons
containing heavy quarks in ultra-relativistic nuclear collisions.

\begin{figure}[hbt]
\includegraphics[width=.48\textwidth]{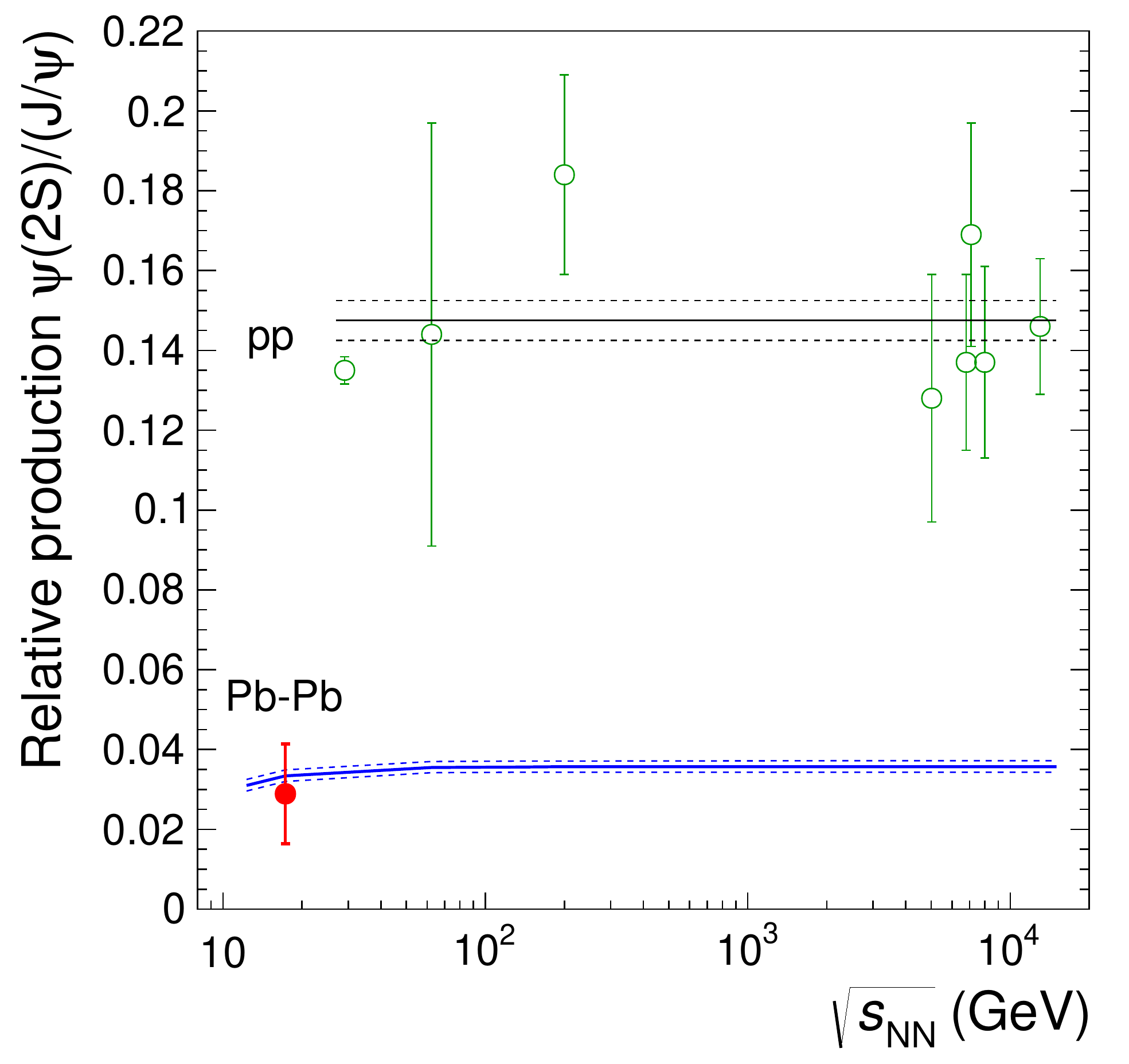}
\caption{Relative production of $\psi(2S)$ and \jpsi
  mesons as a function of collision energy. The data points for proton-proton collisions are from experiments at SPS, HERA, RHIC, and the LHC
  \cite{Abreu:1998rx,Abt:2006va,Adare:2016psx,Aaij:2011jh,Aaij:2012ag,Acharya:2017hjh}.
  The point for central Pb--Pb collisions at SPS energy is from the NA50
  experiment \cite{Alessandro:2006ju}.
The average value of the pp measurements
  is represented by the black horizontal line with dashed
  uncertainties. The blue band denotes statistical model calculations
  for the temperature parameterization from heavy-ion fits, see
  Fig.~\ref{fig:edep}.}
\label{fig:psip2j}
\end{figure}

As a consequence, a very transparent observable to verify the thermal
origin of heavy flavor hadrons produced in a nuclear collision is the
abundance ratio of different resonance states such as $\psi(2S)/({\rm J}/\psi)$
in the charm- or $Y(2S)/Y(1S)$ in the bottom-sector. Indeed, the first
measurement of the $\psi(2S)/(\jpsi)$ abundance ratio at SPS energy
\cite{Alessandro:2006ju} demonstrated that there are clearly different
production mechanisms for charmonia in elementary and nuclear
collisions. This is demonstrated in Fig.~\ref{fig:psip2j}.
While in elementary collisions this ratio is of the order
of 0.15 and hardly varies with energy, the value observed in central
Pb--Pb collisions is more than a factor of four smaller, and is
remarkably consistent with the assumption that these charmonia are
produced at the phase boundary as are all other hadrons.

The chemical freeze-out temperature barely varies with energy beyond
$\sqrt{s_{\rm NN}}=10$ GeV. Since the charm production cross section
drops out in the $\psi(2S)/(\jpsi)$ ratio, the prediction of the
statistical hadronization model for central Pb--Pb collisions at LHC
energy is $\psi(2S)/(\jpsi) =0.035$ with a precision indicated in
Fig.~\ref{fig:psip2j}. Recently, the ALICE collaboration released
the first (\pt-integrated) data  on the above ratio;
the preliminary result is, within the still considerable experimental
uncertainties, well consistent with the statistical hadronization prediction,
lending further support for thermalization of charm quarks in the hot
fireball and the production of charmed hadrons at the phase boundary.

It is also important to assess to which degree the produced charmonia
participate in the hydrodynamic expansion of the fireball. This can be
done by measuring the second Fourier coefficient of the angular
distribution of \jpsi mesons projected onto a plane perpendicular
to the beam direction, the so-called elliptic flow.  Already the first
measurement of \jpsi elliptic flow at the LHC \cite{ALICE:2013xna}
pointed towards rather large values of the elliptic flow
coefficients. The recent measurement at $\sqrt{s_{\rm NN}}=5.02$ TeV from the
ALICE collaboration \cite{Acharya:2017tgv} establishes the detailed flow
pattern as a function of transverse momentum. The large elliptic flow observed
for \jpsi mesons is similar to that observed for open charm mesons \cite{Acharya:2017qps} 
and surprisingly close to the flow coefficients for light hadrons.
This provides strong support for charm quark kinetic thermalization, in
agreement with the statistical hadronization assumption, and implies that
charm quarks couple to the medium in a similar way as light flavor quarks.
Within the current statistical accuracy the \jpsi data at RHIC are
compatible with a null flow signal \cite{Adamczyk:2012pw}.  An elliptic
anisotropy signal was measured for \jpsi mesons at SPS energy
\cite{Prino:2008zz} and was interpreted as a path-length dependence of
the screening.

The response of charmonia produced in ultra-relativistic nuclear
collisions to the medium of the fireball is characterized by the
nuclear modification factor $R_{\mathrm{AA}}$. This factor is
constructed as the ratio between the rapidity densities for \jpsi
mesons produced in nucleus-nucleus (AA) collisions and proton-proton
collisions, scaled by the number of nucleon-nucleon
collisions in a given centrality bin. Clearly, if all charmonia in the final state would
originate from hard scattering processes in the initial state,
$R_{\mathrm{AA}} = 1$.

In the original color screening model \cite{Matsui:1986dk}
$R_{\mathrm{AA}}$ was expected to be significantly reduced from unity,
and to decrease with collision centrality and energy, due to
increasing energy density of the medium. The early experimental
situation until 2009, i.e. before LHC turn-on, is summarized in
\cite{Kluberg:2009wc}. Indeed, the first data from central Pb-Pb
collisions at SPS energy showed a significant suppression which could
be interpreted in terms of nuclear effects and the Debye screening
mechanism. However, the data at RHIC energy exhibited a nearly
identical suppression and an unexpected peaking at mid-rapidity
\cite{Kluberg:2009wc} which could not be reconciled with the
predictions using the color screening model. Both observations on the
energy and rapidity dependence of $R_{\mathrm{AA}}$ for \jpsi
mesons were, however, consistent with their thermal origin
\cite{Andronic:2007bi,BraunMunzinger:2009ih}, albeit with the
qualification of a rather poorly known charm production cross section.

\begin{figure}[hbt]
\includegraphics[width=.5\textwidth]{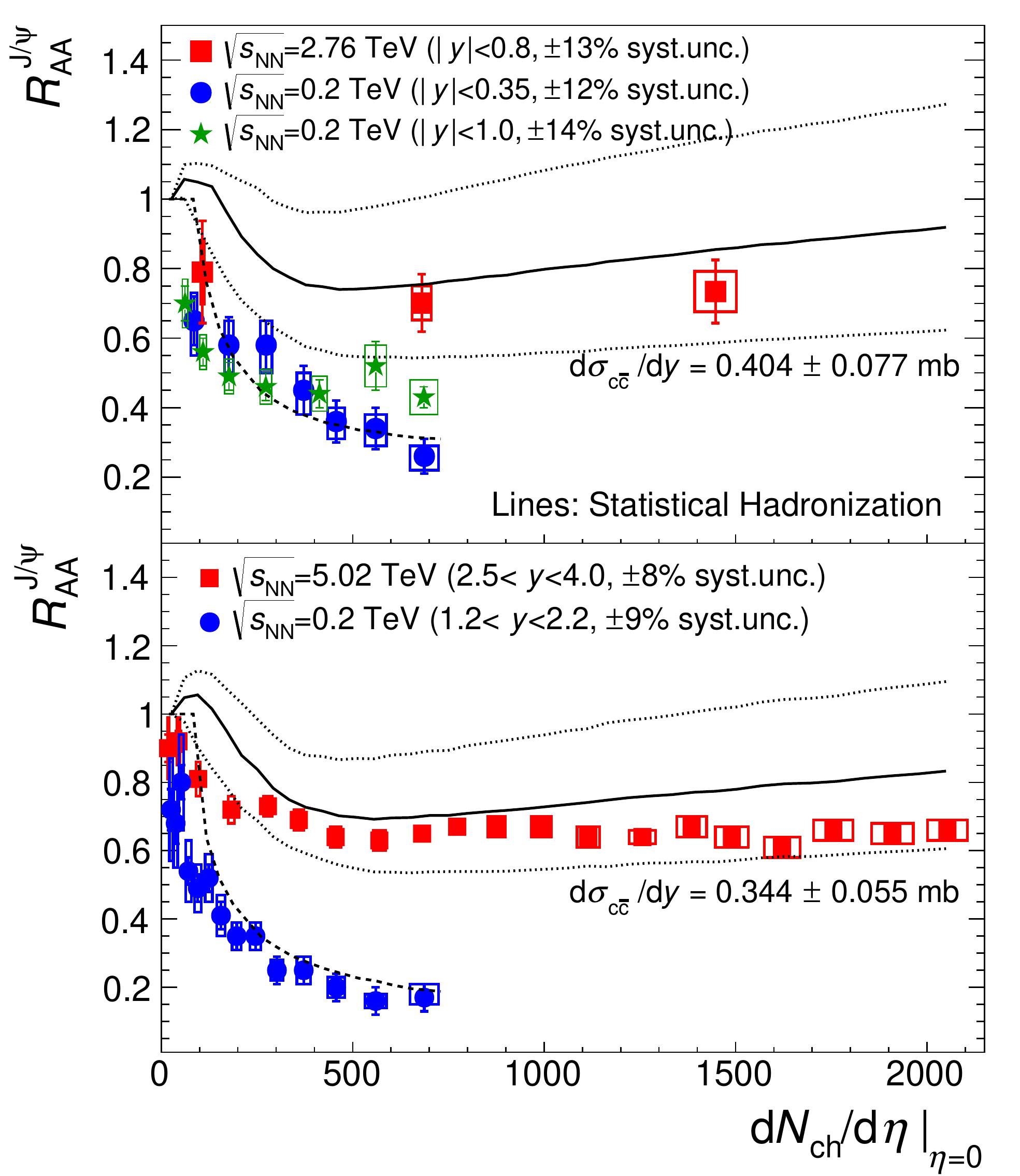}
\caption{The nuclear modification factor
  $R_{\mathrm{AA}}$ for inclusive \jpsi production. The dependence of
  $R_{\mathrm{AA}}$ on the multiplicity density (at $\eta$=0) for midrapidity
  (upper panel) and at forward rapidity (lower panel). The data are for Au--Au
  collisions from the PHENIX collaboration (blue) \cite{Adare:2006ns,Adare:2011yf} and STAR collaboration (green) \cite{Adamczyk:2013tvk} at RHIC
  and for Pb--Pb collisions from the ALICE collaboration  (red) \cite{Abelev:2013ila,Adam:2016rdg} at the LHC.}
\label{fig:raa_jpsi1}
\end{figure}

In the statistical hadronization scenario, the \jpsi nuclear
modification factor $R_{\mathrm{AA}}$ (see above) is obtained by computing the yields in AA collisions while the yields in proton-proton collisions are taken from experimental  data.  The so determined $R_{\mathrm{AA}}$ should increase with increasing collision energy,
implying reduced suppression or even enhancement due to the rapid
increase with energy of the charm production cross section.  Clear
evidence for such a pattern was obtained with the first ALICE
measurements at LHC energy \cite{Abelev:2013ila}. Since then a large
number of additional data including detailed energy, rapidity,
centrality and transverse momentum dependences of $R_{\mathrm{AA}}$
for \jpsi as well as hydrodynamic flow and $\psi(2S)/(\jpsi)$
ratio results have provided a firm basis for the statistical
hadronization scenario \cite{Andronic:2006ky}, with the biggest
uncertainties still related to the not yet measured value of the open
charm cross section in central Pb--Pb collisions. Current results on
\jpsi yields and interpretation within the statistical
hadronization picture are summarized in Fig.~\ref{fig:raa_jpsi1}. A
dramatic increase of $R_{\mathrm{AA}}$ with increasing collision energy is clearly
observed. Furthermore, newer measurements demonstrate, see e.g. Fig. 16 in \cite{Adam:2016rdg}, that the increase is largely concentrated at \jpsi transverse
momentum values less than the mass $m_{\jpsi} = 3.1$ GeV. This latter observation was first predicted in Refs.~\cite{Zhao:2011cv}. 
Both provide further support of the original predictions from the statistical hadronization approach.

\begin{figure}[hbt]
{\includegraphics[width=.48\textwidth]{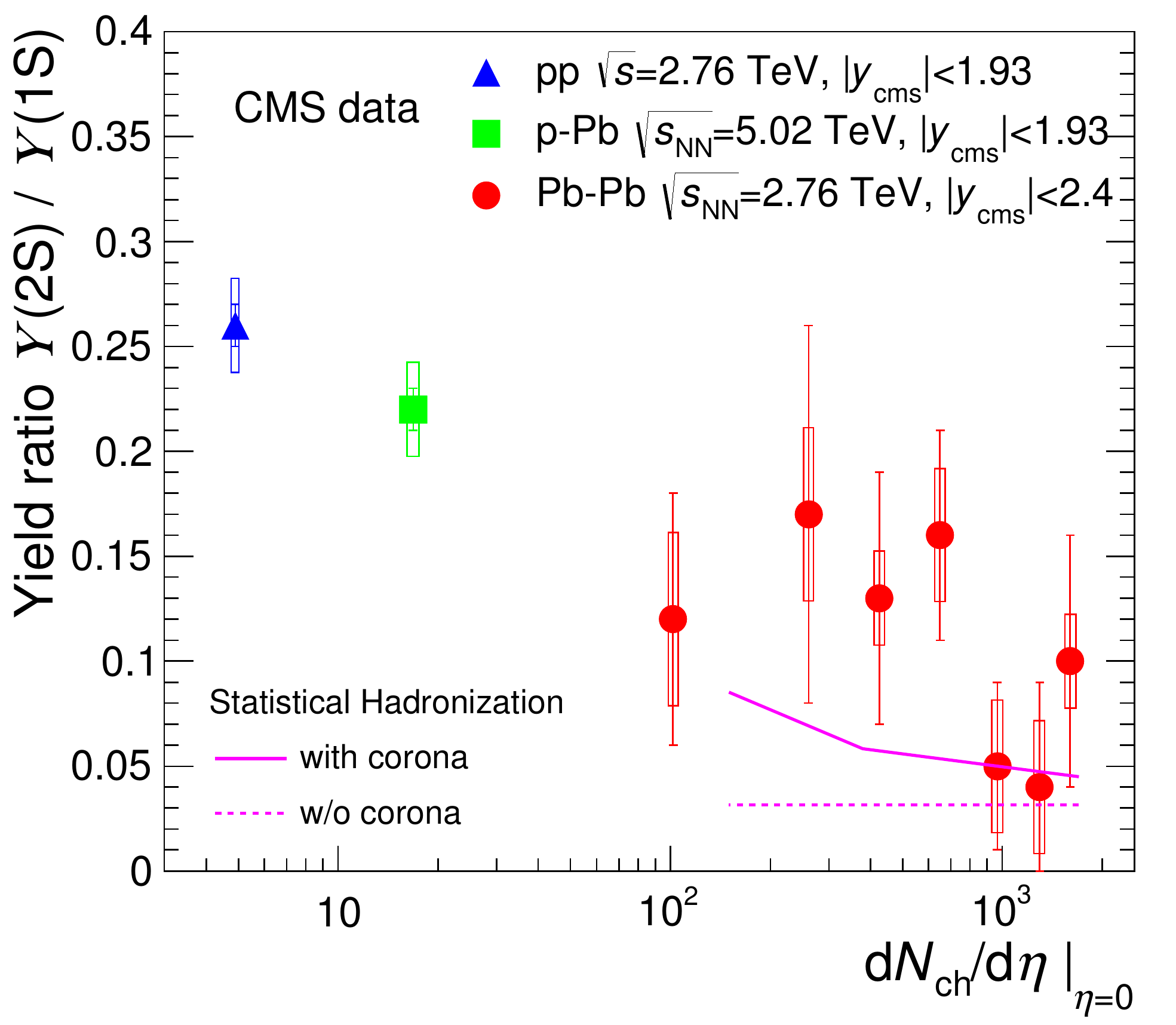}}
\caption{Multiplicity dependence of production ratio  of bottomonium states $\Upsilon(2S)$ and $\Upsilon(1S)$. The data are measured at the LHC in pp, p--Pb and
Pb--Pb collisions \cite{Chatrchyan:2013nza}.
The lines are statistical hadronization predictions for Pb--Pb collisions;
the full line includes an estimate of the contribution of the production in the
corona \cite{Andronic:2006ky} of the colliding nuclei.
}
\label{fig:y}
\end{figure}

Recent measurements of production of bottomonium ($b\bar{b}$) states
at the LHC \cite{Chatrchyan:2012lxa,Chatrchyan:2013nza,Abelev:2014nua}
and at RHIC \cite{Adare:2014hje} can provide further insight into the understanding
of the production dynamics of quarkonia in nuclear collisions.
The nuclear modification factor for the $\Upsilon$ states exhibits at
LHC energies a suppression pattern \cite{Chatrchyan:2012lxa} not unlike that 
expected in the original Debye screening scenario \cite{Emerick:2011xu}.  
On the other hand the observed production ratio $\Upsilon(2S)/\Upsilon(1S)$,
shown in Fig.~\ref{fig:y}, also is consistent with a thermalization
pattern as one approaches central collisions.
Indeed, for central Pb--Pb collisions, this ratio is compatible, with
the value predicted by the statistical hadronization model for $T
\simeq156$ MeV.  This provides the tantalizing possibility of adding
the bottom flavor as an experimental observable to constrain even
further the QCD phase boundary with nucleus-nucleus collision data at
high energies.

An essential ingredient of the statistical hadronization scenario for
heavy quarks is that they can travel, in the QGP, significant
distances to combine with other uncorrelated partons. The observed
increase of the $R_{\mathrm{AA}}$ for $J/\psi$ with increasing
collision energy strongly supports the notion that the mobility of the
heavy quarks is such that it allows travel distances exceeding that of
the typical 1 fm hadronic confinement scale. In fact, for LHC energy,
the volume of a slice of one unit of rapidity of the fireball exceeds
5000 $\rm{fm}^3$, as shown in the previous section, implying that charm quarks 
can travel distances of the order of 10 fm. This results in the possibility 
of bound state formation with all other appropriate partons in the medium with
statistical weights quantified by the characteristics of the hadron
(mass, quantum numbers) at the phase boundary. The results of the
charmonium measurements thereby imply a direct connection to the
deconfinement properties of the strongly interacting medium created in
ultra-relativistic nuclear collisions.

\section*{ Outlook } 
The phenomenological observation of the thermal nature of particle production
in heavy ion collisions at the QCD phase boundary in accord with lattice QCD
raises a number of challenging  theoretical and experimental issues.
An intriguing question is how an isolated quantum
system such as a fireball formed in relativistic nuclear collisions can
reach an apparently equilibrated state. 
Similar questions appear \cite{Rigol:2008aa,Gring1318} in studies of ultra-cold
quantum gases or black holes and may point to a common solution. 
A second area of interest is the mechanism for the formation of loosely bound 
nuclear states in a hot fireball at a temperature exceeding their binding
energies by orders of magnitude. The question of wether there exist
colorless bound states inside a deconfined QGP is related to experimentally
challenging measurements of excited state populations of quarkonia.

Another priority for the field is the direct observation of the restoration of
chiral symmetry and the related critical behavior in relativistic nuclear
collisions with precision measurements and analysis of fluctuation observables.
A highlight would be the observation of a critical endpoint in the QCD phase
diagram.

Making progress with these fundamental issues is at the heart of many ongoing
and future theoretical and experimental investigations.

\vspace{.5cm}
\textbf{Acknowledgements}
K.R. acknowledges partial support by the Extreme Matter Institute EMMI
and the Polish National Science Center NCN under Maestro grant
DEC-2013/10/A/ST2/00106.  This work is part of and supported by the
DFG Collaborative Research Center ''SFB1225/ISOQUANT''.

\bibliography{manuscript1w2}

\end{document}